\documentclass[aps,showpacs,superscriptaddress,preprint]{revtex4}%
\usepackage{amsfonts}
\usepackage{amsmath}
\usepackage{amssymb}
\usepackage{graphicx}%
\providecommand{\U}[1]{\protect\rule{.1in}{.1in}}

\begin{document}
\title{First-principles study of ground state properties of ZrH$_{2}$}
\author{Peng Zhang}
\affiliation{Department of Nuclear Science and Technology, Xi'an
Jiaotong University, Xi'an 710049, People's Republic of
China}\affiliation{LCP, Institute of Applied Physics and
Computational Mathematics, Beijing 100088, People's Republic of
China}
\author{Bao-Tian Wang}
\affiliation{Institute of Theoretical Physics and Department of
Physics, Shanxi University, Taiyuan 030006, People's Republic of
China} \affiliation{LCP, Institute of Applied Physics and
Computational Mathematics, Beijing 100088, People's Republic of
China}
\author{Chao-Hui He}
\affiliation{Department of Nuclear Science and Technology, Xi'an
Jiaotong University, Xi'an 710049, People's Republic of China}
\author{Ping Zhang}
\thanks{Author to whom correspondence should be addressed. E-mail: zhang\_ping@iapcm.ac.cn}
\affiliation{LCP, Institute of Applied Physics and Computational
Mathematics, Beijing 100088, People's Republic of China}
\affiliation{Center for Applied Physics and Technology, Peking
University, Beijing 100871, People's Republic of China}

\begin{abstract}
Structural, mechanical, electronic, and thermodynamic properties of
fluorite and tetragonal phases of ZrH$_{2}$ are systematically
studied by employing the density functional theory within
generalized gradient approximation. The existence of the bistable
structure for ZrH$_{2}$ is mainly due to the tetragonal distortions.
And our calculated lattice constants for the stable face-centered
tetragonal (fct) phase with \emph{c/a}=0.885 are consistent well
with experiments. Through calculating elastic constants, the
mechanically unstable characters of face-centered cubic (fcc) phase
and fct structure with \emph{c/a}=1.111 are predicted. As for
fct0.885 structure, our calculated elastic constants explicitly
indicate that it is mechanically stable. Elastic moduli, Poisson's
ratio, and Debye temperature are derived from elastic constants.
After analyzing total and partial densities of states and valence
electron charge distribution, we conclude that the Zr$-$H bonds in
ZrH$_{2}$ exhibit weak covalent feature. But the ionic property is
evident with about 1.5 electrons transferring from each Zr atom to
H. Phonon spectrum results indicate that fct0.885 and fct1.111
structures are dynamically stable, while the fcc structure is
unstable.
\end{abstract}
 \maketitle

\section{INTRODUCTION}
Many transition metals react readily with hydrogen to form stable
metal hydrides \cite{Fukai}. The common metal hydrides are
technologically attractive materials due to their ability to store
high densities of hydrogen safely. In the case of zirconium
hydrides, another appealing interest arises in nuclear industry,
essentially used as a neutron moderator \cite{Bickel} and fuel rod
cladding materials \cite{Holliger} in nuclear reactors.
Additionally, in the fusion technology applications, zirconium
hydrides are unavoidably precipitated in the excess of the hydrogen
solubility, which may adversely affect several mechanical and
thermal properties in some crystal orientations \cite{Ellis}
involving different stable and metastable hydride phases which may
lead to a significant embrittlement. Furthermore, zirconium hydrides
are also served as a part of the promising new type of actinide
hydride fuels, such as U-(Th-Np-Am)-Zr-H \cite{J.Huang}.
Consequently, great attentions are needed to focus on the basic
scientific research of zirconium hydrides.

Since 1952, there has occurred in literature large amounts of
experimental investigations on zirconium hydrides. The electronic
structures of ZrH$_{x}$ (1.63$\leq$x$\leq$1.94) was studied using
photoelectron spectroscopy and synchrotron radiation \cite{Weaver},
where the Jahn-Teller effect is demonstrated by the changes of
electrons occupation at the Fermi level. An assessed H-Zr phase
diagram was successfully constructed by Zuzek \emph {et al.}
\cite{Zuzek} in 1990. And the transition between the cubic and the
tetragonal phases has been suggested by Ducastelle \emph {et al.}
\cite{Ducastelle} to be of Jahn-Teller type too. From theoretical
point of view, the phase transition between cubic and tetragonal
phases has been investigated by Switendick \cite{Switendick}, where
the non-self-consistent augmented plane wave (APW) method was
employed. Most importantly, the study of the electronic structure
and energetics for the tetragonal distortion in ZrH$_{2}$ has been
extensively performed. Using the pseudopotential method with both
local density approximation (LDA) and general gradient approximation
(GGA), Ackland \cite{Ackland} obtained a double minimum in the
energy surface, one minimum for \emph{c/a}$<$1 and the other for
\emph{c/a}$>$1 with an energy barrier in between of roughly 0.1 eV.
But the ground state for \emph{c/a}$>$1 was in clear disagreement
with the experimental data \cite{Yakel, Bowman1, Bowman2, Bowman3,
Zogal, Niedzwiedz}. Recently, Wolf and Herzig \cite{Wolf} and
Quijano \emph {et al.} \cite{Quijano} studied the accurate total
energy and the electronic structure characteristic of ZrH$_{2}$ by
using LDA and GGA, respectively, within the full potential linear
augmented plane wave (FP-LAPW) method. Both of these works obtained
the correct ground state for \emph{c/a}$<$1 and made a clear
explanation of the cubic to tetragonal distortion.

However, despite the abundant theoretical and experimental research
on zirconium hydrides, relatively little is known regarding their
chemical bonding nature, mechanical properties, and the phonon
dispersions. Until now, the elastic properties, which relate to
various fundamental solid state properties such as interatomic
potentials, equation of states, phonon spectra, and thermodynamical
properties are nearly unkown for ZrH$_{2}$. Moreover, although the
electronic properties as well as the chemical bonding in zirconium
dihydride have been calculated \cite{Wolf}, the study of the bonding
nature of Zr-H involving its mixed ionic/covalent character is still
lacking. As a consequence, these facts inhibit deep understanding of
the zirconium dihydride. Motivated by these observations, in this
paper, we present a first-principles study by calculating the
structural, electronic, mechanical, and thermodynamical properties
of zirconium dihydride. Our calculated results show that the
bistable structure of ZrH$_{2}$ is the fct phase at \emph{c/a}=0.885
and \emph{c/a}=1.111, respectively. The mechanical and dynamical
stability of fct and fcc phases of ZrH$_{2}$ are carefully analyzed.
In addition, through Bader analysis \cite {Bader,Tang} we find that
about 1.5 electrons transfer from each Zr atom to H atom for
ZrH$_{2}$.

\section{Computational method}
Our total energy calculations are self-consistently carried out
using density functional theory (DFT) as implemented in Vienna ab
initio simulation package (VASP) \cite{Kresse}, which is based on
pseudopotentials and plane wave basis functions. All electron
projected augmented wave (PAW) method of Bl\"{o}chl \cite{Blochl1}
is applied in VASP with the frozen core approximation. The
generalized gradient approximation (GGA) introduced by Perdew,
Burke, and Ernzerhof (PBE) \cite{PBE} is employed to evaluate the
electron exchange and correlation potential. Zirconium
4$s$$^{2}$4$p$$^{6}$4$d$$^{2}$5$s$$^{2}$ and hydrogen 1$s$$^{1}$
electrons are treated as valence electrons. The integration over the
Brillouin Zone (BZ) is performed with a grid of special \emph{k}
point-mesh determined according to the Monkhorst-Pack scheme
\cite{Monkhorst}. After convergence test, 12$\times$12$\times$12
\emph{k} point-mesh is chosen to make sure the total energy
difference less than 1 meV per primitive cell. When exploring the
electronic structure, a finer \emph{k} point-mesh
24$\times$24$\times$24 is preferred to generate a high quality
charge density. The expansion of the valence electron plane wave
functions can be truncated at the cutoff energy of 650 eV, which has
been tested to be fully converged with respect to total energy.
Relaxation procedures at ground state are carried out according to
the Methfessel-Paxton scheme \cite{Methfessel} with a smearing width
of 0.1 eV. The geometry relaxation is considered to be completed
when the Hellmann-Feynman forces on all atoms were less than 0.001
eV/\AA. And the accurate total energy calculations are performed by
means of the linear tetrahedron method with the Bl\"{o}chl's
corrections \cite{Blochl2}. The self-consistent convergence of the
total energy calculation is set to 10$^{-6}$ eV.

\section{results}

\subsection{Atomic structure and mechanical properties}

\begin{figure}
\includegraphics[width=6cm,keepaspectratio]{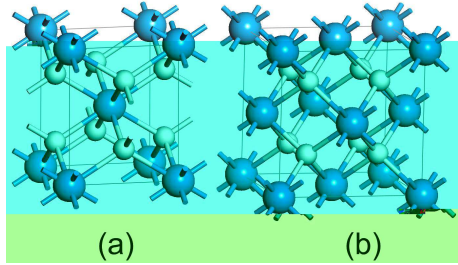}
\caption{\label{fig:epsart}Tetragonal unit cell in space group
$I4/mmm$ (a) and cubic unit cell for ZrH$_{2}$ in space group
$Fm\bar{3}m$ (b) with larger cyan spheres for Zr atoms and the
smaller white H.}
\end{figure}

At ambient condition, the stable zirconium dihydride crystallizes in
a fct structure with space group \emph{I\rm{4}}/\emph{mmm} (No.
139). In its unit cell, there are two ZrH$_{2}$ formula units with
Zr and H atoms in 2\emph{a}:(0,0,0) and
4\emph{d}:(0,$\frac{1}{2}$,$\frac{1}{4}$) sites, respectively [see
Fig. 1(a)]. Each Zr atom is surrounded by eight H atoms forming a
tetragonal and each H connects with four Zr atoms to build a
tetrahedron. The present optimized equilibrium lattice parameters
(\emph{a} and \emph{c}) obtained by fitting the energy-volume data
in the third-order Birch-Murnaghan equation of states (EOS)
\cite{Birch} are 5.030 \AA\  and 4.414 \AA\  (see Table I), in good
agreement with the experimental values. In addition, the fcc
fluorite type structure with space group \emph{Fm$\bar{3}$m} (No.
225) is considered as the metastable phase for ZrH$_{2}$. The cubic
unit cell is composed of four ZrH$_{2}$ formula units with the Zr
and H atoms in 4\emph{a}:(0,0,0) and
8\emph{c}:($\frac{1}{4}$,$\frac{1}{4}$,$\frac{1}{4}$) sites,
respectively [see Fig. 1(b)]. Here, each Zr atom is surrounded by
eight H atoms forming a cube and each H atom connects with four Zr
atoms to build a tetrahedron. Our optimized equilibrium lattice
constant \emph{a} for fcc ZrH$_{2}$ is 4.823 \AA. This value,
although no experimental structural values are available for
comparison, is in good agreement with previous theoretical work in
the literature (see Table I).

\begin{table*}
\caption{Calculated lattice constants, bulk modulus $B_{0}$, and
pressure derivative of the bulk modulus $B_{0}^{'}$ at ground state
from EOS fitting. For comparison, available experimental values and
other theoretical results are also listed.} \label{mechanical}
\begin{tabular}{lcccccc}
\hline\hline
Compounds & Property & Present work & Previous calculation & Experiment\\
\hline
fcc ZrH$_2$ & \emph{a$_{0}$} (\AA)& 4.823 & 4.817$^{\emph{a}}$, 4.804$^{\emph{b}}$& - \\
 & $B_{0}$ (GPa) & 133 & 136$^{\emph{a}}$,152$^{\emph{b}}$ & - \\
 & $B_{0}^{'}$ & 4.01 & - & - \\
fct ZrH$_2$ & \emph{a$_{0}$} (\AA) & 5.030 & 5.021$^{\emph{a}}$, 5.008$^{\emph{b}}$, 5.000$^{\emph{c}}$ & 4.985$^{\emph{d}}$, 4.975$^{\emph{e}}$, 4.982$^{\emph{f}}$ \\
 & \emph{c$_{0}$} (\AA) & 4.414 & 4.432$^{\emph{a}}$, 4.419$^{\emph{b}}$, 4.450$^{\emph{c}}$ & 4.430$^{\emph{d}}$, 4.447$^{\emph{e}}$, 4.449$^{\emph{f}}$ \\
 & $B_{0}$ (GPa) & 127 & - & - \\
 & $B_{0}^{'}$ & 3.62 & - & - \\
\hline\hline
\end{tabular}\\[2pt]
$^{\emph{a}}$ Reference \cite{Quijano}, $^{\emph{b}}$
Reference\cite{Wolf}, $^{\emph{c}}$ Reference \cite{Ackland},
$^{\emph{d}}$ Reference \cite{Yakel}, $^{\emph{e}}$
Reference\cite{Niedzwiedz}, $^{\emph{f}}$ Reference \cite{Bowman1}.
\end{table*}

In present study, the stability of ZrH$_{2}$ has been investigated
by calculating the total energy as a function of \emph{c/a} at the
optimized constant volume of fcc structure, as illustrated in Fig.
2. The energy curve displays two local minima, one with
\emph{c/a}$<$1 (0.885) and the other with \emph{c/a}$>$1 (1.111).
The minimum at \emph{c/a}=0.885 has the lowest energy and
corresponds to fct phase observed experimentally. The fct-fcc
structure energy barrier obtained from this calculation is 0.1048
eV, in good agreement with the Ackland's results \cite{Ackland} but
much larger than the FP-LAPW calculation results
\cite{Wolf,Quijano}.

\begin{figure}
\includegraphics[width=6cm,keepaspectratio]{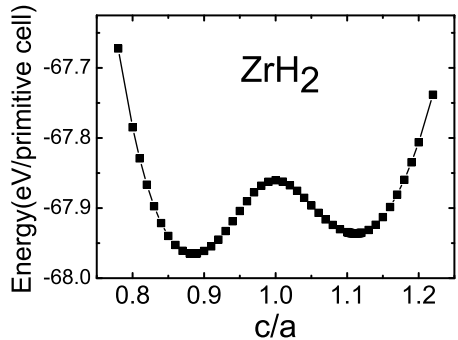}
\caption{\label{fig:epsart}Total energy as a function of the
\emph{c/a} ratio in the fct phase for ZrH$_{2}$.}
\end{figure}

\begin{table*}
\caption{\label{tab:table2}The strain combinations in the strain
tensor [Eq. (1)] to calculate the elastic constants of cubic and
tetragonal ZrH$_{2}$.} \label{strain}
\begin{tabular}{lcccccc}
\hline\hline
Phase & Strain & Parameters (unlisted \emph{e$_{i}$}=0)& \emph{$\Delta$E/V$_{0}$} in \emph{O($\delta$$^{2}$)}\\
\hline fcc  &
\emph{$\epsilon$$^{1}$} & \emph{e$_{1}$}=\emph{e$_{2}$}=\emph{e$_{3}$}=$\delta$ & $(\frac{3}{2}C_{11}+3C_{12})\delta^{2}$\\
& \emph{$\epsilon$$^{2}$}& \emph{e$_{1}$}=\emph{e$_{2}$}=$\delta$ & $(C_{11}+C_{12})\delta^{2}$\\
& \emph{$\epsilon$$^{3}$}& \emph{e$_{4}$}=\emph{e$_{5}$}=\emph{e$_{6}$}=$\delta$ & $\frac{3}{2}C_{44}\delta^{2}$\\
fct  &
\emph{$\epsilon$$^{1}$} & \emph{e$_{1}$}=\emph{e$_{2}$}=\emph{e$_{3}$}=$\delta$ & $(C_{11}+C_{12}+2C_{13}+\frac{1}{2}C_{33})\delta^{2}$\\
& \emph{$\epsilon$$^{2}$}& \emph{e$_{1}$}=\emph{e$_{3}$}=$\delta$ & $\frac{1}{2}(C_{11}+2C_{13}+C_{33})\delta^{2}$\\
& \emph{$\epsilon$$^{3}$}& \emph{e$_{1}$}=\emph{e$_{2}$}=$\delta$ & $(C_{11}+C_{12})\delta^{2}$\\
& \emph{$\epsilon$$^{4}$} & \emph{e$_{3}$}=$\delta$ & $\frac{1}{2}C_{33}\delta^{2}$\\
& \emph{$\epsilon$$^{5}$}& \emph{e$_{4}$}=\emph{e$_{5}$}=$\delta$ & $C_{44}\delta^{2}$\\
& \emph{$\epsilon$$^{6}$}& \emph{e$_{6}$}=$\delta$ & $\frac{1}{2}C_{66}\delta^{2}$\\
\hline\hline
\end{tabular}\\[2pt]
\end{table*}

Elastic constants can measure the resistance and mechanical features
of crystal to external stress or pressure, thus may evaluate the
stability of crystals against elastic deformation. For small strain
$\varepsilon$, Hooke's law is valid. Therefore, we can enforce it
onto the equilibrium lattice, determine the resulting change in the
total energy, and from this information deduce the elastic
constants. The crystal total energy $E(V,\epsilon)$ can be expanded
as a Taylor series \cite{Nye1},
\begin{eqnarray}
E(V,\epsilon)=E(V_{0},0)+V_{0}\sum_{i=1}^{6}\sigma_{i}e_{i}+\frac{V_{0}}%
{2}\sum_{i,j=1}^{6}C_{ij}e_{i}e_{j}+O(\{e_{i}^{3}\}),
\end{eqnarray}
where $E(V_{0},0)$ is the energy of the unstrained system with the
equilibrium volume $V_{0}$, $\epsilon$ is the strain tensor which
has matrix elements $\varepsilon_{ij}$ ($i,j$=1, 2, and 3) defined
by
\begin{eqnarray}
\varepsilon_{ij}=
\left(
\begin{array}
[c]{ccc}%
e_{1} & \frac{e_{6}}{2} & \frac{e_{5}}{2}\\
\frac{e_{6}}{2} & e_{2} & \frac{e_{4}}{2}\\
\frac{e_{5}}{2} & \frac{e_{4}}{2} & e_{3}%
\end{array}
\right),
\end{eqnarray}
and $C_{ij}$ are the elastic constants. Such a strain transforms the
three lattice vectors defining the unstrained Bravais lattice
\{\textbf{a}$_{k}$, \emph{k}=1, 2, and3\} to the strained vectors
$\left\{\mathbf{a}_{k}^{\prime}\right\}$ \cite{Alou} as defined by
\begin{eqnarray}
\mathbf{a}_{k}^{\prime}=(\mathbf{I}+\mathbf{\varepsilon})\mathbf{a}_{k},
\end{eqnarray}
where $\mathbf{I}$ is a unit $3\times3$ matrix. Each lattice vector
$\mathbf{a}_{k}$ or $\mathbf{a}_{k}^{\prime}$ is a $3\times1$
matrix.

For the cubic phase, there are three independent elastic constants,
i.e., $C_{11}$, $C_{12}$, and $C_{44}$, which can be calculated
through a proper choice of the set of strains $\{e_{i},i=1,...,6\}$
(see Table II). As for the fct phase, the six independent elastic
constants, i.e., $C_{11}$, $C_{12}$, $C_{13}$, $C_{33}$, $C_{44}$,
and $C_{66}$, can be obtained from six different strains listed in
Table II. To avoid the influence of high order terms on the
estimated elastic constants, we have used very small strains, i.e.
within $\pm$1.5\%. Thus, in our calculations the strain amplitude
$\delta$ is varied in steps of 0.006 from $\delta$=$-$0.036 to
0.036. And the total energies $E(V,\delta)$ at these strain steps
are calculated and fitted through the strains with the corresponding
parabolic equations of $\Delta E/V_{0}$ to yield the required
second-order elastic constants.

Our calculated elastic constants for fcc ZrH$_{2}$ and fct ZrH$_{2}$
at \emph{c/a}=0.885 and \emph{c/a}=1.111 are collected in Table III.
Obviously, fct ZrH$_{2}$ at \emph{c/a}=0.885 is mechanically stable
due to the fact that its elastic constants satisfy the following
mechanical stability criteria \cite{Nye2} of tetragonal structure:
\begin{eqnarray}
&C_{11}>0, C_{33}>0, C_{44}>0, C_{66}>0, \nonumber\\
&(C_{11}-C_{12})>0, (C_{11}+C_{33}-2C_{13})>0, \nonumber\\
&[2(C_{11}+C_{12})+C_{33}+4C_{13}]>0.
\end{eqnarray}
Nevertheless, the fct phase at \emph{c/a}=1.111 is mechanically
unstable in light of the fact that $C_{11}$ is smaller than
$C_{12}$. As for the fcc structure of ZrH$_{2}$, it is also
mechanically unstable because of the fact that its elastic constants
cannot meet the following mechanical stability criteria \cite{Nye2}
of cubic structure:
\begin{eqnarray}
&C_{11}>0, C_{44}>0, C_{11}>|C_{12}|, (C_{11}+2C_{12})>0.
\end{eqnarray}

\begin{table*}
\caption{\label{tab:table3}Calculated elastic constants in units of
GPa for ZrH$_{2}$ at zero pressure.} \label{elastic}
\begin{tabular}{lcccccccccccccccc}
\hline\hline
Phase & \emph{C$_{11}$} & \emph{C$_{12}$} &\emph{C$_{44}$} & \emph{C$_{13}$} & \emph{C$_{33}$} & \emph{C$_{66}$}\\
\hline
fcc & 82.6 & 159.7 & -19.5 \\
fct0.885 & 165.6 & 140.9 & 30.5 & 106.8 & 145.5 & 60.6 \\
fct1.111 & 125.7 & 145.5 & 30.9 & 115.0 & 190.6 & 42.0 \\
\hline\hline
\end{tabular}\\[2pt]
\end{table*}

\begin{table*}
\caption{\label{tab:table4}Calculated bulk modulus, shear modulus,
Young's modulus, Poisson's ratio, density, transverse sound
velocity, longitudinal sound velocity, average sound velocity, and
Debye temperature of the fct phase of ZrH$_2$ at \emph{c/a}=0.885.}
\label{mechanical0885}
\begin{tabular}{lcccccccccc}
\hline\hline \emph{B}(GPa) & \emph{G}(GPa) & \emph{E}(GPa) &
$\upsilon$ & $\rho$(g/cm$^{3}$)
& \emph{v$_{t}$}(km/s) & \emph{v$_{l}$}(km/s) & \emph{v$_{m}$}(km/s) & $\theta_{D}$(K) \\
\hline
130 & 29 & 80 & 0.397 & 5.5201 & 2.2823 & 5.5197 & 2.5825 & 364.9\\
\hline\hline
\end{tabular}\\[2pt]
\end{table*}

After obtaining elastic constants, bulk modulus $B$ and shear
modulus $G$ for fct phase of ZrH$_{2}$ at \emph{c/a}=0.885 can be
calculated from the Voigt-Reuss-Hill (VRH) approximations
\cite{Voigt,Reuss,Hill}. Then the Young's modulus \emph{E} and
Poisson's ratio $\upsilon$ are calculated through $E=9BG/(3B+G)$ and
$\upsilon=(3B-2G)/[2(3B+G)]$. The Debye temperature ($\theta_{D}$)
is obtained using the relation \cite{Anderson}
\begin{eqnarray}
\theta_{D}=\frac{h}{k_{B}}\left[\frac{3n}{4\pi}\left(\frac{N_{A}\rho}{M}\right)\right]^{1/3}v_{m},
\end{eqnarray}
where $h$ and $k_{B}$ are Planck's and Boltzmann's constants,
respectively, $N_{A}$ is Avogadro's number, $\rho$ is the density,
$M$ is the molecular weight, $n$ is the number of atoms in the
molecule, and $v_{m}$ is the average wave velocity. The average wave
velocity in the polycrystalline materials is approximately given as
\begin{eqnarray}
v_{m}=\left[\frac{1}{3}(\frac{2}{v_{t}^{3}}+\frac{1}{v_{l}^{3}})\right]^{-1/3},
\end{eqnarray}
where $v_{t}$=$\sqrt{G/\rho}$ and $v_{l}$=$\sqrt{(3B+4G)/3\rho}$ are
the transverse and longitudinal elastic wave velocity of the
polycrystalline materials, respectively.

All the calculated results for fct0.885 are collected in Table IV.
Results of fcc and fct1.111 can not be obtained due to their
mechanically unstable nature. Note that we have also calculated the
bulk modulus $B$ by EOS fitting. The derived bulk modulus for
fct0.885 turns out to be very close to the one obtained from the
above VRH approximation, which again indicates that our calculations
are consistant and reliable. It is well known that the shear modulus
$G$ represents the resistance to plastic deformation, while the bulk
modulus $B$ can represent the resistance to fracture. A high (low)
$B/G$ value is responsible for the ductility (brittleness) of
polycrystalline materials. The critical value to separate ductile
and brittle materials is about 1.75. Using the calculated values of
bulk modulus $B$ and shear modulus $G$ for fct0.885, the $B/G$ value
of 4.48 can be obtained. For element Zr, the $B/G$ value of 2.63 can
be derived from the elastic data in Ref. \cite{Liu}. Therefore, fct
phase of ZrH$_{2}$ is rather ductile and its ductility is more
predominant than that of element Zr. As for the Poisson's ratio, the
value is well within the range from 0.25 to 0.45 for typical metals.
Unfortunately, no reliable experimental and theoretical results
concerning on the mechanical properties in the literature are
available for comparison. We hope that our calculated elastic
constants and elastic moduli can be illustrative in the realistic
application of the mechanical data for ZrH$_{2}$.

\begin{figure*}
\includegraphics[width=16cm,keepaspectratio]{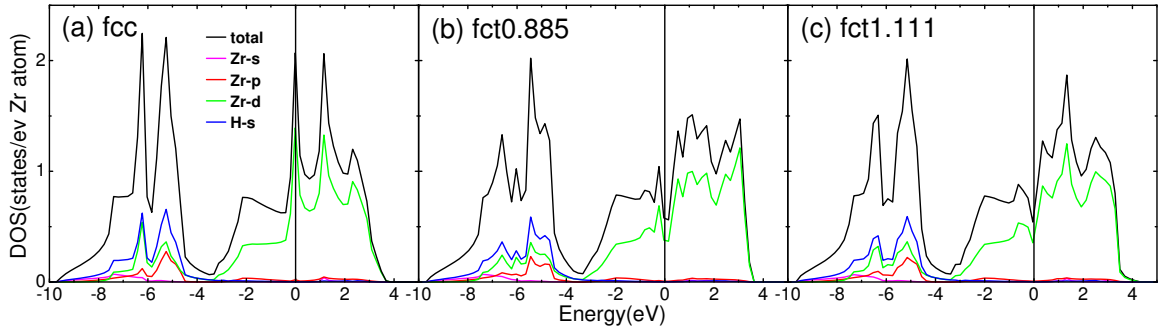}
\caption{\label{fig:epsart}Total and orbital-resolved local
densities of states for (a) fcc, (b) fct0.885, and (c) fct1.111
structures of ZrH$_{2}$. The Fermi energy level is set at zero.}
\end{figure*}

\subsection{Electronic structure and charge distribution}
Basically, all the macroscopical properties of materials, such as
hardness, elasticity, and conductivity, originate from their
electronic structure properties as well as the nature of the
chemical bonding. Therefore, it is necessary to perform the
electronic structure analysis of ZrH$_{2}$. The calculated total
densities of states (DOS) and the orbital-resolved partial densities
of states (PDOS) of fcc ZrH$_{2}$, fct ZrH$_{2}$ at \emph{c/a}=0.885
and \emph{c/a}=1.111, are shown in Fig. 3. Wholly, the occupation
properties of ZrH$_{2}$ in fct and fcc phases are similar. The
pseudogap below the Fermi level, located at around $-$3.5 eV, for
all three systems can be observed. The energy region below the
pseudogap has predominant feature of hybridization for H $1s$
orbital and Zr $4d$ and Zr $4p$ orbitals. States in this region have
critical contribution to the Zr-H and H-H bonding state. However,
from the pseudogap to the Fermi level, H $1s$ states contribute very
little to the total DOS, while the Zr $4d$ states dominate in this
energy range. The fact that the DOS occupation has obvious sharp
peak at the Fermi level for fcc ZrH$_{2}$ implies that the cubic
phase should be unstable. In contrast, a strong reduction in the DOS
at the Fermi level is observed in the two fct structures. This
evident difference between fcc and fct phase in the vicinity of the
Fermi level is caused by the splitting of the band in this region.
As a consequence, the Fermi level in the two tetragonal phases is
not located at a peak of the DOS but at a local minimum, which leads
to a reduction in the electronic contribution to the total energy.
This behavior has been referred to as a Jahn-Teller mechanism, as
indicated by previous experiments and theoretical works
\cite{Weaver,Bowman1,Wolf,Quijano}.

\begin{figure*}
\includegraphics[width=16cm,keepaspectratio]{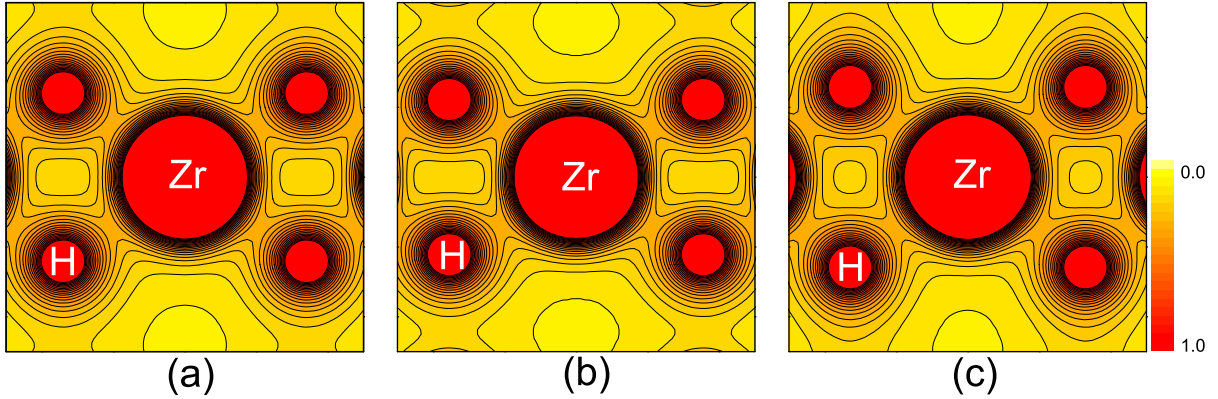}
\caption{\label{fig:epsart}Valence charge density of (a) fcc phase
in (110) plane, (b) fct0.885 in (100) plane, and (c) fct1.111 in
(100) plane. The contour lines are drawn from 0.0 to 1.0 at 0.05
e/{\AA}$^{3}$ intervals.}
\end{figure*}

To analyze the ionic/covalent character of zirconium dihydride, in
the following we will carefully investigate the valence charge
density distribution. The calculated valence charge density maps of
the fct ZrH$_{2}$ at \emph{c/a}=0.885 and \emph{c/a}=1.111 in (100)
plane and fcc ZrH$_{2}$ in (110) plane are plotted in Fig. 4.
Obviously, the charge densities around Zr and H ions are all near
spherical distribution with slightly deformed towards the direction
to their nearest neighboring atoms. Covalent bridges, which
represent the bonding nature of Zr$-$H bonds, is clearly shown in
Fig. 4. In Fig. 5, we plot the line charge density distribution
along the nearest Zr$-$H bonds. Clearly, three line charge density
curves vary in almost the same way along the Zr$-$H bonds. One can
find a minimum value of charge density for each bond at around the
bridge locus (indicated by the arrow in Fig. 5). These minimum
values are listed in Table V. Although these values are much smaller
than 0.7 e/\AA\  for Si covalent bond, they are prominently higher
than 0.05 e/\AA\  for Na$-$Cl bond in typical ionic crystal NaCl.
Therefore, there are weak but clear weights in Zr$-$H bonds in both
fcc and fct phases of ZrH$_{2}$.

\begin{figure}
\includegraphics[width=0.5\linewidth]{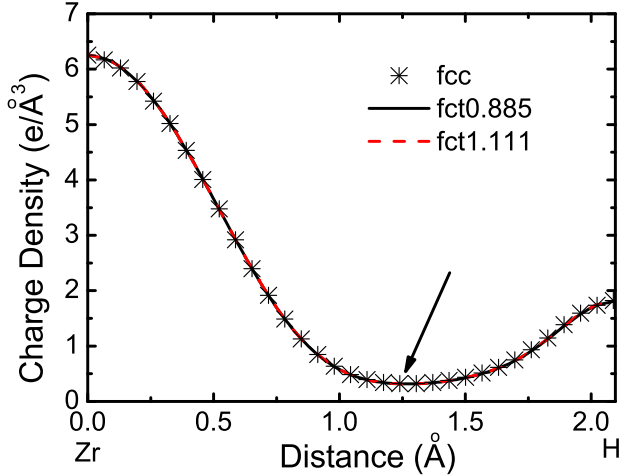}
\caption{\label{fig:epsart}The line charge density distribution
between Zr atom and the nearest neighbor H atom for fcc, fct0.885,
and fct1.111 structures of ZrH$_{2}$.}
\end{figure}

In addition, we have performed the Bader analysis \cite {Bader,Tang}
for the three typical ZrH$_{2}$ cells. The charge ($Q_{B}$) enclosed
within the Bader volume ($V_{B}$) is a good approximation to the
total electronic charge of an atom. In this work, the default charge
density grids for one primitive cell are 56$\times$56$\times$56,
56$\times$56$\times$56, 60$\times$60$\times$60 for fcc, fct0.885,
and fct1.111 cells, respectively. To check the precision, the charge
density distributions are calculated with a series of $n$ times
finer grids ($n$=2,3,4,5,6,7). The deviation of the effective charge
between the six and the seven times finer grids is less than 0.2\%.
Thus we perform the charge density calculations using the seven
times finer grid (392$\times$392$\times$392,
392$\times$392$\times$392, 420$\times$420$\times$420) for fcc,
fct0.885, and fct1.111 cells, respectively. The calculated results
are presented in Table V. Note that although we have included the
core charge in charge density calculations, only the valence charge
are listed since we do not expect variations as far as the trends
are concerned. From Table V, we find that the Bader charges and
volumes for fcc and fct ZrH$_{2}$ are almost equal to each other.
This shows similar ionic character, through a flux of charge (about
1.5 electrons for each Zr atom) from cations towards anions, for
zirconium dihydride in their stable phase and metastable phases. The
slight difference in charge distribution is mainly due to the
structure distortion. Thus the ionicity of Zr-H bonds for ZrH$_{2}$
is also evident either for its fcc structure or fct structures.

\begin{table*}
\caption{\label{tab:table5}Calculated charges \emph{Q$_{B}$} and
volumes \emph{V$_{B}$}(\AA$^{3}$) according to Bader partitioning as
well as the Zr$-$H distance (\AA) and relevant minimum values of
charge densities (e/\AA$^{3}$) along the Zr$-$H bonds for
ZrH$_{2}$.} \label{bader}
\begin{tabular}{lcccccccccc}
\hline\hline
Compounds & \emph{Q$_{B}$}(Zr) & \emph{Q$_{B}$}(H) & \emph{V$_{B}$}(Zr) & \emph{V$_{B}$}(H) & Zr-H distance & Charge density$_{min}$\\
\hline
fcc & 10.484 & 1.758 & 14.809 & 6.618 & 2.088 & 0.320 \\
fct0.885 & 10.512 & 1.744 & 14.953 & 6.544 & 2.095 & 0.316 \\
fct1.111 & 10.514 & 1.743 & 14.954 & 6.545 & 2.094 & 0.317 \\
\hline\hline
\end{tabular}\\[2pt]
\end{table*}

\subsection{Phonon dispersion curves}
Phonon frequencies of the crystalline structure is one of the basic
aspects when considering the phase stability, phase transformations,
and thermodynamics of crystalline materials. Employing the
Hellmann-Feynman theorem \cite{Nielsen1,Nielsen2} and the direct
method \cite{Parlinski}, we have calculated the phonon curves along
some high symmetry directions in the BZ, together with the phonon
DOS. For the phonon dispersion calculation, we use the
2$\times$2$\times$2 fcc (fct) supercell containing 96 (48) atoms for
fcc (fct) ZrH$_{2}$ and the 4$\times$4$\times$4 Monkhorst-Pack
$k$-point mesh for the BZ integration. In order to calculate the
Hellmann-Feynman forces, we displace four and eight atoms,
respectively, for fcc and fct ZrH$_{2}$ from their equilibrium
positions and the amplitude of all the displacements is 0.03\AA. The
calculated phonon dispersion curves along the
$\Gamma$-$X$-$K$-$\Gamma$-$L$-$X$-$W$-$L$ directions for fcc
ZrH$_{2}$ and along the $\Gamma$-$N$-$P$-$X$-$\Gamma$-$Z$ directions
for fct ZrH$_{2}$  are displayed in Fig. 6. For both fcc and fct
ZrH$_{2}$, there are only three atoms in their primitive cell.
Therefore, nine phonon modes exist in the dispersion relations. Due
to the fact that zirconium is much heavier than hydrogen, the
vibration frequency of zirconium atoms is apparently lower than that
of hydrogen atoms. As a result, evident gap between the optic modes
and the acoustic branches exits and the phonon DOS of fcc (fct)
ZrH$_{2}$ can be viewed as two parts. One is the part lower than 7.4
(7.1) THz, where the main contribution comes from the zirconium
sublattice, while the other part range from 30.7 to 37.9 (29.6 to
38.5) THz is dominated by the dynamics of the light hydrogen atoms.
As shown by Fig. 6(b) and 6(c), all frequencies are positive for the
two fct structures. This assures that the two fct phases are both
dynamically stable. In contrast, one can see from Fig. 6(a) that the
fcc phase of ZrH$_{2}$ is dynamically unstable. The transverse
acoustic (TA) mode close to $\Gamma$ point becomes imaginary along
the $\Gamma$-$K$ and $\Gamma$-$L$ directions. The minimum of the TA
branch locates at the $\Gamma$-$K$ direction. Therefore, the fcc to
fct phase transition probably can occur along the [101] direction.
This kind of dynamical instability of fcc phase compared to the
stable fct phase is well consistent with our above-discussed
mechanical stability analysis of ZrH$_{2}$.

\begin{figure*}
\includegraphics[width=1.0\linewidth]{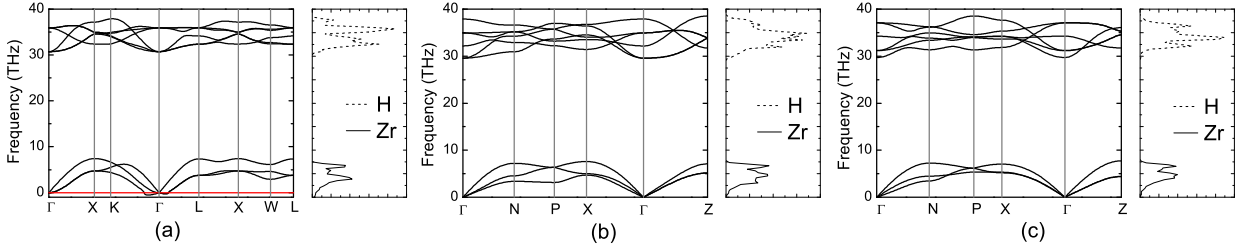}
\caption{\label{fig:epsart}Calculated phonon dispersion curves along
the high-symmetry directions (left panel) and the corresponding PDOS
(right panel) at ground state for (a) fcc ZrH$_{2}$, (b) fct
ZrH$_{2}$ at \emph{c/a}=0.885, and (c) fct ZrH$_{2}$ at
\emph{c/a}=1.111.}
\end{figure*}

\section{CONCLUSION}
In summary, we have investigated the structural, mechanical,
electronic, and thermodynamic properties of ZrH$_{2}$ in its stable
and metastable phases by means of first-principles DFT-GGA method.
Our optimized structural parameters are well consistent with
previous experiments and theoretical calculations. Tetragonal
distortion leads to bistable structures with \emph{c/a}$>$1 and
\emph{c/a}$<$1, the latter of which corresponds to the
experimentally observed low-temperature phase. Our total energy
calculations illustrate that the most stable phase of ZrH$_{2}$ is
the fct structure with \emph{c/a}=0.885. Elastic constants, various
moduli, Poisson's ratio, and Debye temperature are calculated for
fcc and two fct phases of ZrH$_{2}$. Mechanically unstable nature
for fcc and fct1.111 structures are predicted. For fct0.885,
mechanical stability and dynamical stability are firstly predicted
by our elastic constants results and phonon dispersion calculation.
The occupation characters of electronic orbital also accord well
with experiments. Through Bader analysis, we have found that the
Zr$-$H bonds exhibit weak covalent character while the ionic
property is dominant with about 1.5 electrons transferring from each
Zr atom to H. Our calculated phonon curves of fcc ZrH$_{2}$ have
shown that the TA mode becomes imaginary close to $\Gamma$ point
along the [101] direction.

\section{ACKNOWLEDGMENTS}
This work was partially supported by NSFC under Grants No. 90921003
and No. 60776063.

\end{document}